\documentclass[%
 reprint,
amsmath,amssymb,
 aps,pra,
]{revtex4-2}
\usepackage{adjustbox}
\usepackage{graphicx}
\usepackage{dcolumn}
\usepackage{subfigure}
\usepackage{float}
\usepackage{url}
\usepackage{orcidlink}
\usepackage{hyperref}

\begin{document}

\preprint{APS/123-QED}

\title{A Comprehensive Convolutional Neural Network Architecture Design using Magnetic Skyrmion and Domain Wall}

\author{Saumya Gupta$^{1}$\orcidlink{0009-0001-7636-1553}}
\author{Venkatesh Vadde$^{1}$\orcidlink{0009-0001-4874-7563}}%
\author{Bhaskaran Muralidharan$^{1}$\orcidlink{0000-0003-3541-5102}}%
\author{Abhishek Sharma$^{2}$\orcidlink{0000-0002-3635-3081} }%
\email{ abhishek@iitrpr.ac.in}
\affiliation{$^{1}$Department of Electrical Engineering, Indian Institute of Technology Bombay, Powai, Mumbai-400076, India
}
 \affiliation{$^{2}$Department of Electrical Engineering, Indian Institute of Technology Ropar, Rupnagar, Punjab-140001, India}%

\date{\today}
\begin{abstract}
Spintronic-based neuromorphic hardware offers high-density and rapid data processing at nanoscale lengths by leveraging magnetic configurations like skyrmion and domain walls. Here, we present the maximal hardware implementation of a convolutional neural network (CNN) based on a compact multi-bit skyrmion-based synapse and a hybrid CMOS domain wall-based circuit for activation and max-pooling functionalities. We demonstrate the micromagnetic design and operation of a circular bilayer skyrmion system mimicking a scalable artificial synapse, demonstrated up to 6-bit (64 states) with an ultra-low energy consumption of 0.87~fJ per state update.~We further show that the synaptic weight modulation is achieved by the perpendicular current interaction with the labyrinth-maze like uniaxial anisotropy profile, inducing skyrmionic gyration, thereby enabling long-term potentiation (LTP) and long-term depression (LTD) operations.  Furthermore, we present a simultaneous rectified linear (ReLU) activation and max pooling circuitry featuring a SOT-based domain wall ReLU with a power consumption of 4.73~$\mu$W. The ReLU function, stabilized by a parabolic uniaxial anisotropy profile, encodes domain wall positions into continuous resistance states coupled with the HSPICE circuit simulator. Our integrated skyrmion and domain wall-based spintronic hardware achieves 98.07\% accuracy in convolutional neural network (CNN) based pattern recognition task, consuming 110~mW per image.
\end{abstract}

\maketitle


\section{\label{Introduction}Introduction}
\vspace{-4mm}
{A}{rtificial} intelligence\cite{nature2018big} simulates human cognition, demanding substantial resources to handle the exponential data growth. 
In contrast, neuromorphic computing\cite{furber2016large} emulates the human brain structure and functionality adapting to more complex tasks, and achieving low power consumption.
The ability to train an artificial neural network (ANN) model with minimal energy consumption is one of the key features of neuromorphic computing over Von-Neumann architecture. While, the conventional Moore approach, relies on CMOS technology, exploring beyond Moore by investigating spintronics-based neuromorphic\cite{grollier2020neuromorphic} circuitry presents a promising alternative for low-power devices and systems. The various emerging spintronic-based devices ranging from synapses to leaky integrate and fire (LIF) neurons\cite{liu2022magnetic}
~using the magnetic tunnel junctions(MTJs)\cite{cai2023unconventional}, domain walls(DW)\cite{desai2022chip} and magnetic skyrmions\cite{song2020skyrmion} have been proposed for neuromorphic computing\cite{jadaun2022adaptive}.\\
\indent In the class of ANN, the convolutional neural networks (CNN)\cite{zhang2021time} inspired by the visual cortex, form the foundation of many complex image recognition and generation architectures such as GAN\cite{goodfellow2020generative}, Unet, and ResNet\cite{zhang2017image}.
The CNN\cite{lecun1995convolutional} is a deep learning model that captures the spatial and temporal dependencies in data using the feature extraction and fully connected stage for image classification, detection and segmentation applications\cite{langkvist2016classification}. The feature extraction involves multiple layers such as the convolutional layer, activation layer and pooling layer. While the fully connected stage classifies the image based on learned features. Rapid processing and reduced energy consumption of spintronic devices can be leveraged for hardware implementation of convolutional layers using magnetic skyrmion based synapse, and DW based device for activation and pooling layers.\\
\indent Magnetic skyrmions, two-dimensional solitons with local spin whirls in magnetic materials \cite{fert2017magnetic}, have attracted significant interest for in-memory\cite{sisodia2022robust} computation. 
Skyrmions offer advantages such as small size, ultra-low depinning current density, and particle-like characteristics\cite{nagaosa2013topological}, with the ability to accumulate within a compact area due to their topological protection and stability. These features make skyrmions suitable for synapstic devices in neuromorphic computing, where they can extract features from input data by utilizing change in weights.\\
\indent The spin-orbit torque (SOT)-driven DWs are also widely explored for synapse applications \cite{bhowmik2019chip}\cite{siddiqui2019magnetic}\cite{liu2021domain} due to the non-volatility offered by the DW's position. However, the requirement of long-term stability, smaller current operation, and high synaptic resolution (see methodology) position the magnetic skyrmion as a more suitable candidate for its application as a synapse. 
Also, the persistent DW motion even after the current is turned off hinders the realization of a practical synapse design based on DW. Whereas, skyrmion's stable configuration\cite{wang2022fundamental} renders them immune to the persistent velocity issue observed in DWs.
Traditionally, in a skyrmion-based synapse\cite{das2023bilayer,huang2017magnetic}, accommodating more states in an orderly manner requires increased device length, raising concerns about shape anisotropy and multiple domain formation. However, the robustness of skyrmions with respect to compact packing in a circular geometry allows for better state resolution through multiple sensing MTJs\cite{zhao2024electrical}, enabling smaller device sizes compared to DWs with a large number of states. It can be noted that CNN requires a large number of multi-state\cite{lee2018deep} synapses. Hence, our proposed design favours skyrmions over domain walls for synaptic operation due to the scalability, translating to a linear and symmetric weight update.\\
\indent The CNN implementation also involves an activation (ReLU: rectified linear unit) and max-pooling function. While the activation function facilitates complex data learning by introducing non-linearity in the system, the max-pooling function selects the maximum element of the feature map, pooling all the prime features from the image.
The ReLU function rectifies the vanishing gradient descent problem in deep learning models providing an advantage over other activation functions such as sigmoid and hyperbolic tangent function. In our proposed design,  we have implemented the domain wall-based activation function due to their faster reset (few ns) and the ability to provide continuous resistance states~(see Sec.\ref{res rel maxpool}) over magnetic skyrmion.\\
\indent Most spintronic-based neural networks feature a segment of the network as their key spintronics component, complemented by software-based activation and pooling functions\cite{yadav2023demonstration}\cite{song2020skyrmion}\cite{ojha2024neuromorphic}\cite{han2024neuromorphic}\cite{desai2022chip}\cite{he2017tunable}, which makes it difficult to gauge the full potential of the maximal spintronic network. Hence, we present the design of a multi-state ordered skyrmionic synapse along with DW-based activation and max-pooling function through a hybrid CMOS-micro-magnetic simulation platform. Our multi-state skyrmionic  synapse utilizes a circular geometry to alleviate the possible concern about the shape anisotropy\cite{cowburn2000property}.~Whereas, the simultaneous activation and Max pooling circuit design uses a DW with a modified uniaxial anisotropy profile to address the concern of the persistent DW motion~(see sec. \ref{Domain wall-based ReLU device})
Finally, we implement a maximal spintronic CNN architecture with a crossbar array using the proposed skyrmionic synapse (convolution) and hybrid CMOS DW-based circuit performing ReLU-Max Pooling functionalities simultaneously. 

\begin{figure}[t]
    \centerline{\includegraphics[width=1\columnwidth]{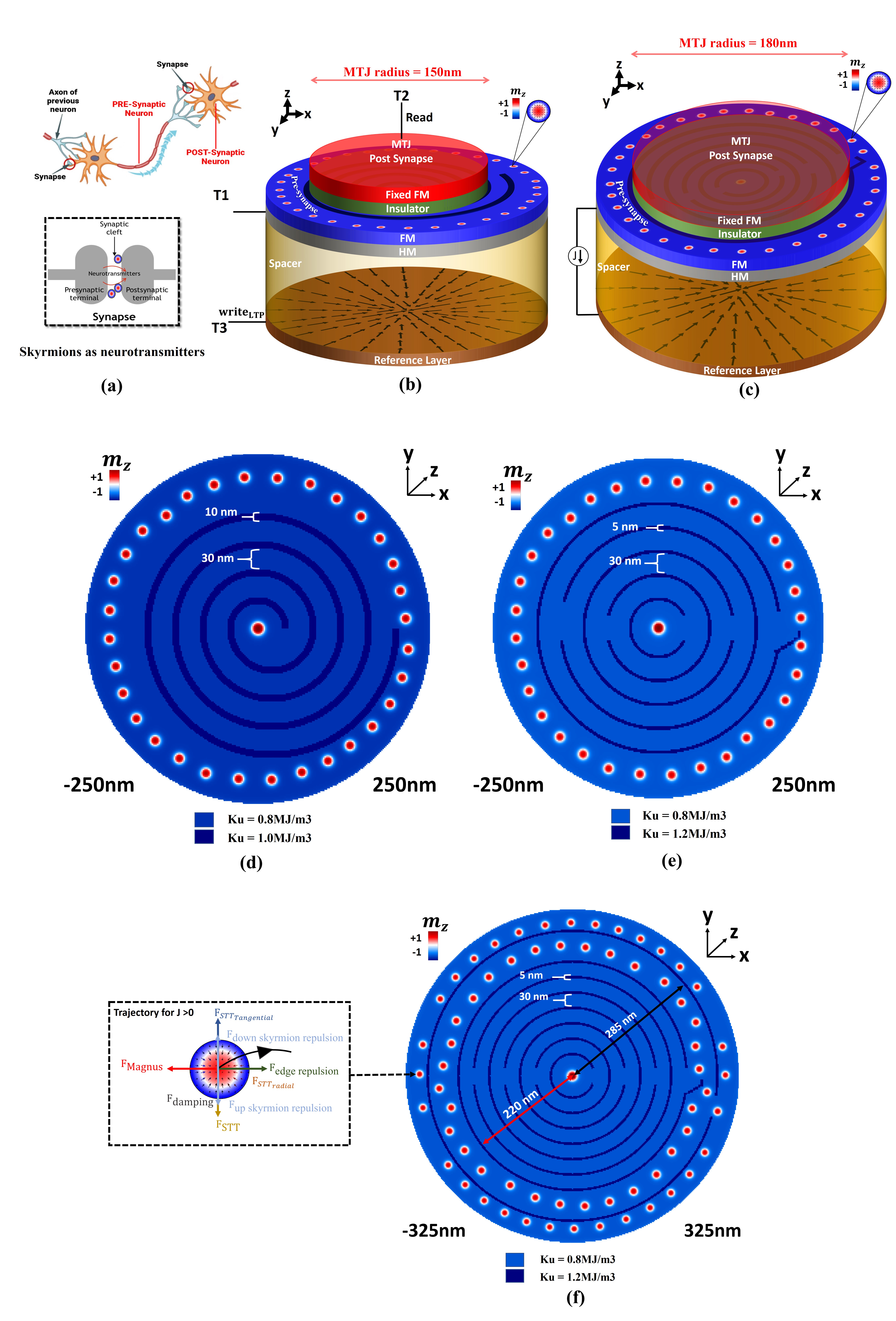}}
    \caption{(a)~Schematic of a biological neuron connected via a synapse.
    A 3D schematic of the proposed magnetic skyrmion-based 5-bit artificial Synapse (b) with spiral uniaxial anisotropy profile. (c) with concentric uniaxial anisotropy profile.
    The top view of a skyrmionic synapse device shows 32 skyrmions around the periphery and one skyrmion pinned at the center (d) with spiral-like uniform and (e) concentric-like uniform uniaxial anisotropy profile. (f) A 6-bit skyrmionic synapse device having 64 skyrmions with a concentric-like uniform uniaxial anisotropy profile. The magnetization direction is color-coded: blue is into the plane, white is in the plane, and red is out of the plane.}
    \label{Fig.1}
\end{figure}
\section{\label{Device Model}Results}
\vspace{-4mm}
\begin{figure}[t]
\centerline{\includegraphics[width=0.85\columnwidth]{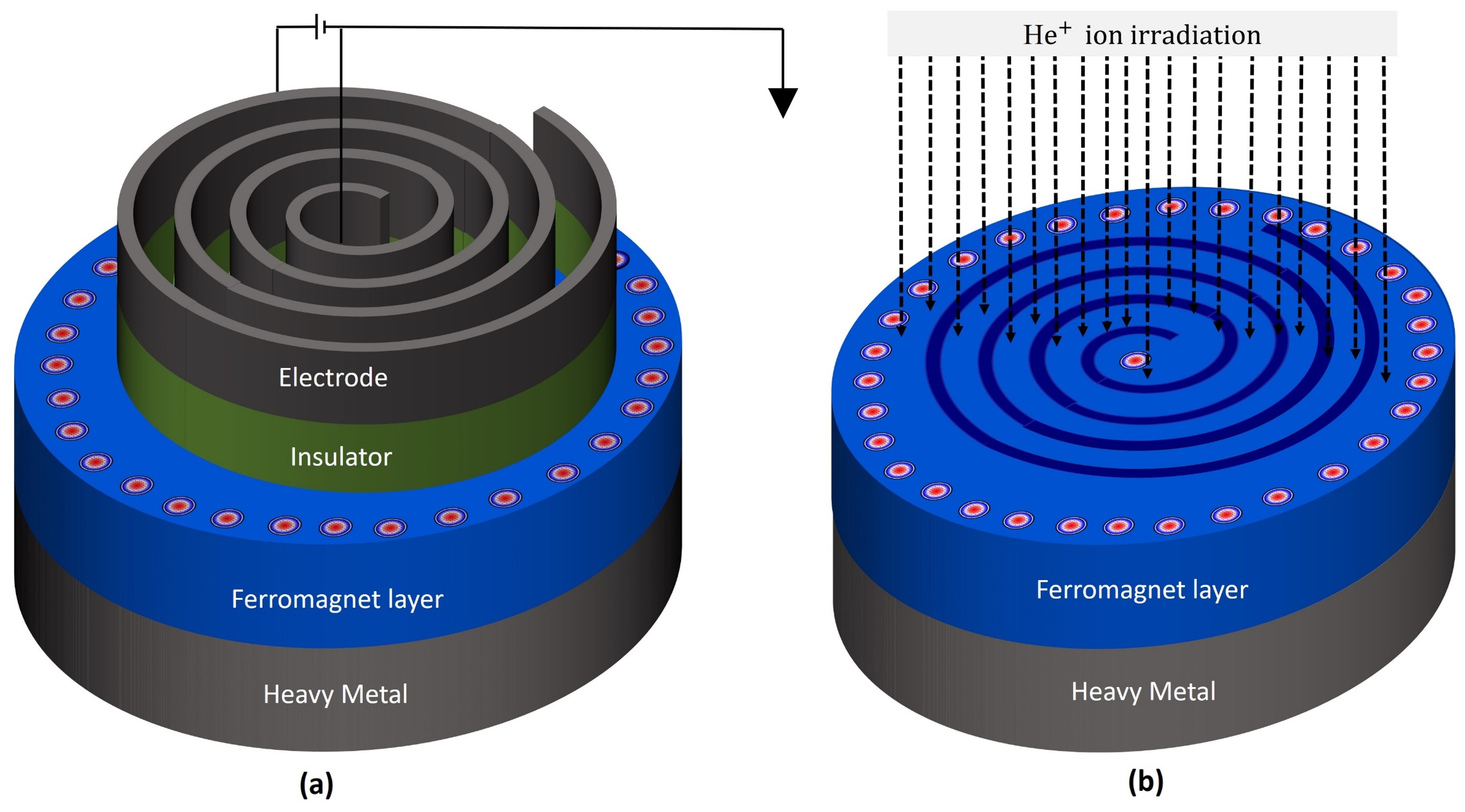}}
\caption{Different ways to create uniaxial anisotropy profiles in the device. (a)~Schematic of the desired PMA profile in the device created via inducing voltage on the electrode gate through an insulator. (b)~$He^+$ ion irradiation technique modifies ultrathin film's magnetic properties (nanometer scale) using a focused ion beam to pattern the required magnetic tracks.}
\label{Fig.2}
\end{figure}
\subsection{\label{Skyrmionic synapse device}Design of Skyrmion based synapse device}
\vspace{-4mm}
\indent The biological neural network comprises neurons interconnected through synapses, as depicted in Fig.~\ref{Fig.1}(a) synapse typically consists of three components: a pre-synaptic region, a synaptic cleft, and a postsynaptic region. In our case, skyrmions function as neurotransmitters, being stored and released by the pre-synaptic region, while the postsynaptic region receives them. During synaptic transmission, an electrical pulse triggers the skyrmions to move individually from the pre-synaptic region to the postsynaptic region, facilitated by the synaptic cleft: a narrow gap between the pre-and postsynaptic regions. Applying a positive current pulse causes skyrmions to travel from the pre-synaptic region to the postsynaptic region through the synaptic cleft, leading to long term potentiation(LTP). Conversely, a negative current pulse induces skyrmions to move from the postsynaptic region back to the pre-synaptic region via the synaptic cleft, resulting in long term depression(LTD). \\
\indent The proposed skyrmion based synapse device comprises of two circular FM layers separated by an insulator, namely the ferromagnet layer(FM) and the reference layer(RL). The FM layer is in contact with a heavy metal (HM) layer at the bottom, giving rise to the emergence of Dzyaloshinskii-Moriya interaction at the HM/FM interface. The FM layer has a perpendicular magnetic anisotropy(PMA). The reference layer's magnetization is fixed, having a vortex-like spin polarization as shown in Fig.~\ref{Fig.1}(b),  which is consistent with the experimental observations of vortex reference layer~\cite{phatak2012direct, wintz2013topology}.~A circular detector (MTJ) region of reduced radius~(see Fig.~\ref{Fig.1}(b,c)) is formed by sandwiching the oxide layer between the FM and the free FM layer~(m\textsubscript{z}: red ,opposite magnetization to the free FM layer), forms the post-synaptic region of our synapse device. The area between the circumference of the nanodisk and the detector forms the pre-synaptic region where the skyrmions are nucleated in a stable form such that the skyrmion-skyrmion repulsion aids in the skyrmion motion on current application in the device, analogous to the biological model as shown in Fig.~\ref{Fig.1}(a).\\
\indent We present design of four synapse device based on skyrmion having a labyrinth maze-like uniaxial anisotropy profile (unicursal: one path that leads you from the entrance to the center) that forms the nano track for our proposed devices with two different designs: a spiral-like uniaxial anisotropy profile for a 5-bit (32+1 skyrmions) synapse, as shown in Fig.~\ref{Fig.1}(d) and three concentric-like uniaxial anisotropy profiles for a 4-bit (16+1 skyrmions), 5-bit (32+1 skyrmions)~(see Fig.~\ref{Fig.1}(e)), and 6-bit (64+1 skyrmions)~(see Fig.~\ref{Fig.1}(f)) synapse~respectively.
\begin{table*}
\caption{The synapse device parameters}
\begin{ruledtabular}
    \begin{adjustbox}{max width=\textwidth}
    \small
          \begin{tabular}{ccccccccc}
          \textbf{Bit} &~ \textbf{No. of states} & \textbf{Profile} & \textbf{Dimension} & $\bf{R_{dev}}$ & $\bf{R_{det}}$  & \bf{K\textsubscript{u}} $\bf{ track_{sp}}$ & $\bf{K\textsubscript{u}}~\bf{track_{th}}$ &$\bf{K\textsubscript{u}_{high}}$\\ \hline
     \\
        4 & 16 & Concentric & 330nm x 330nm x 0.5nm & 165nm & 110nm & 20nm & 5nm &1.0 MJ/$m^3$\\
 
        5 & 32 & Spiral& 500nm x 500nm x 0.5nm& 250nm & 150nm & 30nm & 10nm & 1.0 MJ/$m^3$\\
 
         5 & 32 & Concentric& 500nm x 500nm x 0.5nm& 250nm & 180nm & 30nm & 5nm & 1.2 MJ/$m^3$\\
 
         6 & 64 & Concentric& 650nm x 650nm x 0.5nm& 325nm & 220nm& 30nm & 5nm& 1.2 MJ/$m^3$\\
         \end{tabular}
\end{adjustbox}
\end{ruledtabular}
\label{table1}
\end{table*}
The device description of our proposed devices is as follows: In a 32-state, 5-bit skyrmionic synapse based on a spiral uniaxial anisotropy~(K\textsubscript{u}) profile, with a radius of 250~nm and thickness of 0.5~nm. The device radius is 250~nm while the detector region is of radius 150~nm. The spiral K\textsubscript{u} track passage is 30~nm wide with a 10~nm thickness band having a high uniaxial anisotropy value of $\mathrm{1~MJ/m^{3}}$~(dark blue region) falling well within the fabrication capabilities\cite{yang2023magnetic, zhao2016direct}.\\
\indent Similarly, for 16, 32 and 64-state skyrmionic synapses based on a concentric uniaxial anisotropy profile, the device dimensions along with device radius~(R\textsubscript{dev}), detector radius~(R\textsubscript{det}), K\textsubscript{u} track passage~(K\textsubscript{u}~track\textsubscript{sp}), K\textsubscript{u} track passage thickness~(K\textsubscript{u}~track\textsubscript{th}) with their corresponding high uniaxial anisotropy values~(K\textsubscript{u}\textsubscript{high}) ~are shown in~Table~\ref{table1}. Additionally, all the devices have a constant uniaxial anisotropy value of $\mathrm{0.8~MJ/m^{3}}$~(light blue region), where the constricted high K\textsubscript{u}~(darker region) passage is formed.\\
\indent Figure~\ref{Fig.1} shows the locally injected orderly skyrmions in the FM layer of the device , demonstrating configurations with 16 states, 32 states, and scalable up to 64 states~(Video~\cite{Videolink}). Skyrmions in thin film and multi-layers experience a transverse deflection subjected to a gyrotropic force when driven by the spin current called the skyrmion hall effect(SkHE).~Our device design benefits from the Skyrmion hall effect(SkHE), which is usually disregarded due to annihilation issues.\\
\indent To move the skyrmions, a current perpendicular to the plane (CPP) generates a spin-polarized current that makes the skyrmions excited to a steady state of persistent circular trajectory but is directed towards the constricted spiral or concentric-like profile in unison along the boundary created by high K\textsubscript{u} nano-track followed by a trail of skyrmions. At the spiral or concentric opening the skyrmions not only experience edge repulsion but also a self-directional repulsive force that guides the trail of skyrmions towards the nano track created by the boundaried high K\textsubscript{u} tracks or rings.\\
The applied spin-polarised current results in spin torque acting on the FM layer, where the spin polarisation takes the form of a vortex-like state from the in-plane~(`write' current: Terminal T3 to T1) magnetized reference layer. The vortex-like magnetic configuration with a radial magnetization configuration is set at a polarization angle of $180^{o}$ (\textit{e.g.},$\Psi$  =$180^{o}$) \cite{phatak2012direct,wintz2013topology} which simulates the polarization vector through the relation m\textsubscript{RL} = (cos$\Phi_{p}$, sin $\Phi_{p}$, 0), where $\Phi_{p}$ = tan$^{-1} y/x + \Psi $ and (x y) are spatial coordinates in the film plane with origin being at the center of the nanodisk. \\
\indent When an electrical current is injected along the +z-direction into the system, the spin current drives the skyrmion into a clockwise gyration towards the center but as it encounters the high K\textsubscript{u} passage, and starts to move along the track until the duration of the applied current.~However, at the center of the disk, it encounters an already pinned skyrmion which does not let the incoming skyrmions get pinned at the center but repels them. Now, the current pulse is switched along –z-direction and the skyrmions are driven towards the edge again following the same constricted high K\textsubscript{u} path and eventually reaching a steady state of anti-clockwise gyration along the edge of the disk. This orderly controlled motion of the skyrmions in the nano-disk can be used as a synapse.\\
The area where these high K\textsubscript{u} constricted tracks are created is used as a detector region forming the post-synaptic region.
 The applied current density with spin polarisation (according to the $\mathrm{m_{RL}}$) injected in the FM varies the magnetization of the free layer of the MTJ~(detector)\cite{li2022experimental} as the trail of skyrmions reach the detector region. This can be electrically readout using magnetoresistance\cite{huang2017magnetic}, along the `read' path between terminal T2 to T1. A range of conductance values corresponding to the applied current pulse train is obtained as synaptic weights. The conductance value obtained from magnetization is described in \ref{electrical readout}.~When no skyrmions are present under the MTJ, the conductance value corresponds to G\textsubscript{min}, the lowest conductance. When all the skyrmions move under the MTJ, the conductance value corresponds to G\textsubscript{max}, the maximum conductance.
 ~Figure~\ref{fig.5}(a,b) shows a linear~relation between the vertical axis representing normalized conductance or weights~(see \ref{electrical readout})~versus the horizontal axis representing the number of skyrmions, which is an effect of the magnetic moment alignment of the center of the skyrmion with the fixed FM layer of the detector. We investigated our devices using identical current pulses with spin polarisation according to the $\mathrm{m_{RL}}$ applied into the FM based on the artificial synapse operations. The synaptic operation consists of initiation, potentiation(LTP), and depression(LTD) analogous to the persistent strengthening \textit{i.e,} the increasing number of skyrmions during the positive current pulse or weakening \textit{i.e.,} decreasing number of skyrmions at a negative current pulse of synaptic connections showcasing the LTP and LTD synapse operations respectively as plotted and marked in Fig.~\ref{fig.5}(c, d, e)~with the vertical axis representing the weights and the horizontal axis as the pulse number. The synapse can also have short-term plasticity~(STP) which is a constant state between the LTP and LTD slopes or short-term change that decays over tens of ms or minutes. It is analogous to the maximum conductance value achieved after all the skyrmions have reached the detector. Beyond this point, even with the continued current application, the conductance change remains constant~(see Fig.\ref{fig.5}(d,e)) unless the current pulse train is reversed, leading to LTD operation.\\
\indent The skyrmions placement and motion in an orderly arrangement allow for better resolution of weights \textit{i.e.,} the system can accurately represent and differentiate between a greater number of discrete weight values or states. Additionally, a higher magnitude of resolution can be achieved by using multiple MTJ corresponding to each skyrmion.\\
It can be noted that the current density magnitude exceeds $\mathrm{2.5~MA/cm^{2}}$, few skyrmions can be annihilated either at the center or the edge of the nano-disk due to an increased Magnus force with the increasing current. Still, a linear profile is maintained even at an increased current density.~Also, the center pinned skyrmion if not present then the incoming skyrmions near the center are pushed by the current in the high K\textsubscript{u} region, which in turn provides a repulsive force resulting in reduced skyrmion size and hence, the skyrmion vanishes. Thus, the center pinning of skyrmion  holds importance in device design.\\
\indent The spiral or concentric-shaped constricted K\textsubscript{u} profiles can be obtained using He\textsuperscript{+} ion irradiation techniques\cite{balan2023improving,juge2021helium} which provide the desired profile without getting into the complexity of voltage-controlled magnetic anisotropy (VCMA)~\cite{xia2018skyrmion} methods for modifying anisotropy as shown in Fig.~\ref{Fig.2}(a,b). The high uniaxial anisotropy track thickness and value are chosen such that the skyrmions don't crossover them but instead align along the narrow tracks. The concentric K\textsubscript{u} profile devices have openings along 180\textsuperscript{o} with a slightly lower opening at the start having an inclined L-shaped wedge to avoid skyrmion dodging at the entry supporting the skyrmion trajectory.\\
\indent  In the 5-bit~(concentric-like K\textsubscript{u} profile) and 5-bit~ (spiral and concentric-like profile) skyrmionic synapse, the skyrmions are placed in the pre-synaptic region along the circumference, see snapshot in Fig.~\ref{Fig.1}(d,e). Additionally, the 6 bit~(concentric-like K\textsubscript{u} profile) synapse device has a two-line concentric formation of the skyrmions along the circumference separated by an additional high K\textsubscript{u} ring of value 1.2~$\mathrm{MJ/m^{3}}$ and 5nm thickness.~The introduced high K\textsubscript{u} outer ring (radius: 285~nm (Fig.~\ref{Fig.1}(f)))~enables more skyrmion accommodation in a diameter of 650~nm with 64 skyrmions compared to the inner diameter of 500~nm accommodating 32 skyrmions. If we place the 64 skyrmions in a concentric formation without the outer separation ring, the outer skyrmions will push the inner skyrmions toward the high K\textsubscript{u} regions under the detector. That can cause skyrmion annihilation due to the push and pull experienced by the presence of high K\textsubscript{u} repulsion, edge repulsion, and neighboring skyrmion repulsion resulting in reduced size and finally vanishing under the influence of applied current. The inner ring skyrmions enter the post-synaptic region followed by the outer ring skyrmions.
The spiral-like profile has an opening on the right side along 180\textsuperscript{o} supporting the skyrmion trajectory. If the skyrmions reach the innermost track they might annihilate in the high K\textsubscript{u} region near or at the center due to the experienced forces. The more preferred Uniaxial anisotropy profile is the one with concentric K\textsubscript{u} rings because of two reasons: 1) The skyrmion coming in the innermost ring can circle or move around the center-pinned skyrmion without getting annihilated.~2)~The total duration of device operation is reduced, as explained in \ref{res syn conc}.\\
\indent The energy consumption required to move the skyrmions in the nano tracks for hardware implementation in the neural network are done as per Eq.~(\ref{Eenergy}), described in \ref{Simulation methods}. For a charge current of 2.1~mA and pulse period~(T\textsubscript{p}) of 2~ns corresponds to $\mathrm{E_{write}}$ = 0.8724~fJ/state for the 4-bit skyrmionic synapse having a concentric K\textsubscript{u} profile.~For a charge current of 4.9~mA and T\textsubscript{p} of 2~ns corresponds to $\mathrm{E_{write}}$ = 2.0028~fJ/state for both 5-bit skyrmionic synapses having a concentric and spiral K\textsubscript{u} profile.~For a charge current of 8.3~mA and T\textsubscript{p} of 2.5~ns corresponds to $\mathrm{E_{write}}$ = 4.2309~fJ/state for 6-bit skyrmionic synapses having a concentric K\textsubscript{u} profile.\\
In the next subsections, a detailed working of each profile with the applied pulse train is studied.
\subsection{\label{res synap spiral}Synapse based on spiral Ku profile}
\begin{figure}
\centerline{\includegraphics[width=1\columnwidth]{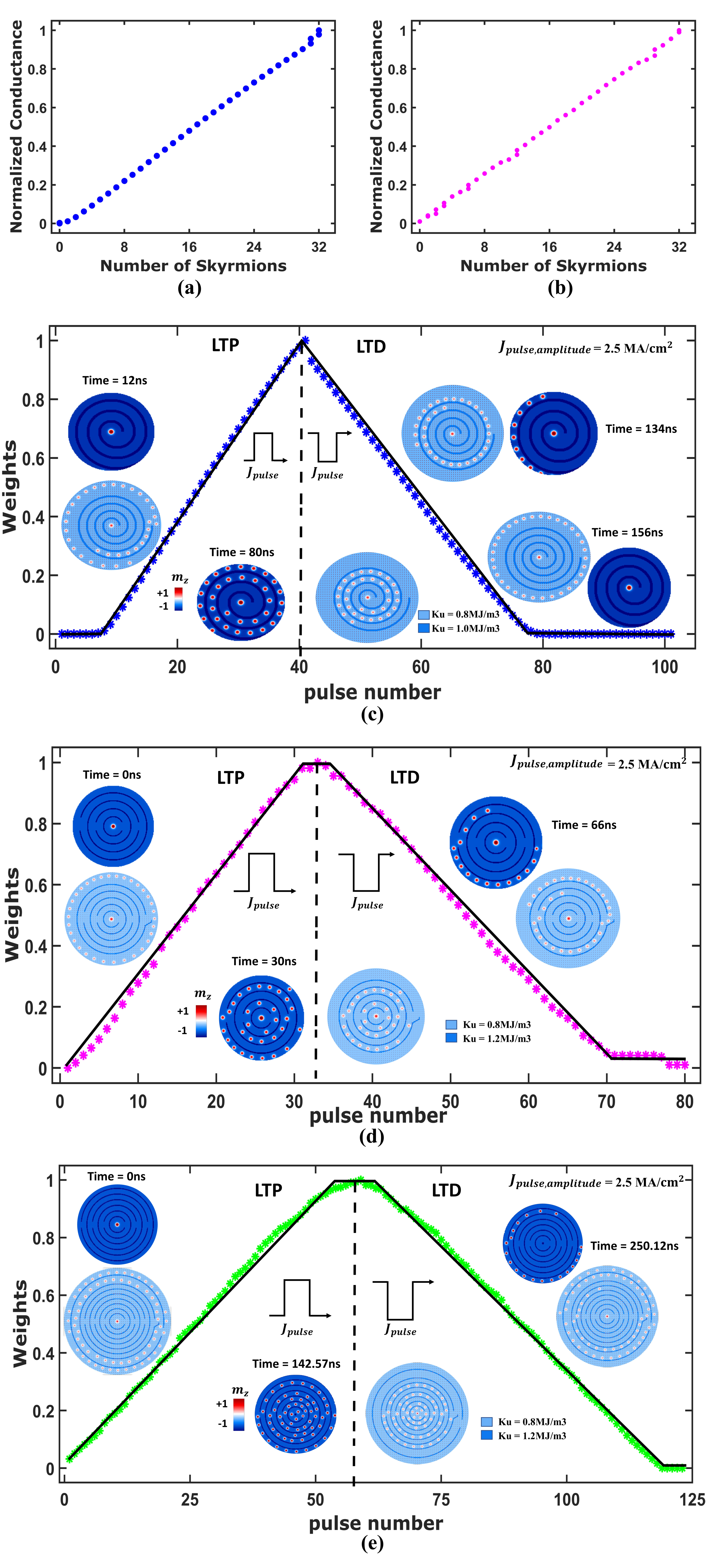}}
\caption{Normalized conductance versus the number of skyrmions in the post-synaptic region (a) A 5-bit synapse device with spiral uniaxial anisotropy profile. (b) A 5-bit synapse device with a concentric-like uniaxial anisotropy profile.
Synapse weight versus pulse number during LTP and LTD with snapshots of micro-magnetic simulations showing m\textsubscript{z}, for the full device and the post-synaptic region at different time stamps (c) A 5-bit skyrmionic synapse with spiral uniaxial anisotropy profile (d) A 5-bit skyrmionic synapse with concentric uniaxial anisotropy profile. (e) 6-bit skyrmionic synapse with concentric uniaxial anisotropy profile.}
\label{fig.5}
\end{figure}

\vspace{-4mm}
After the skyrmions are locally injected and relaxed for 5 bit synapse (spiral-like K\textsubscript{u} profile) as shown in Fig.~\ref{fig.5}(c), full device snapshots (light blue) with color-coded K\textsubscript{u} and  m\textsubscript{z} are shown at different simulation time stamps along with the post-synaptic or detector region of the device (in dark blue color). The current pulse of value J = 2.5~MA/cm\textsuperscript{2} at a pulse width of 1.5~ns and period of 2~ns is applied for a total duration of 186.15~ns.\\
\indent In the initiation process, the skyrmions are brought near the detector region for a duration of 16~ns. The snapshot in Fig.~\ref{fig.5}(c) shows no skyrmion under the detector at t = 12~ns but several skyrmions can be seen inside the spiral opening which is not a part of the detector region. The detector starts at 150~nm from the center.\\
\indent For the LTP operation, a positive pulse train starts moving the skyrmion in the clockwise direction and the skyrmions start to enter the detector region one by one at each current pulse. The LTP operation takes place from 16~ns to 80~ns, where we observe a linear increase in weights, with G\textsubscript{max} corresponding to all the skyrmions under the detector region as shown in the snapshot at t = 80~ns. \\
\indent For the LTD operation, a negative current pulse train starts moving the skyrmion in an anti-clockwise direction forcing skyrmions to move out of the spiral, corresponding to each pulse. Thus, linearly decreasing the weights as shown in the snapshot at t = 134~ns with 8+1 skyrmions under the detector region. The LTD operation takes place from 80~ns to 154~ns. After which the device takes 154~ns to 186~ns (pulse number 93) for a complete reset. As shown at time t = 156~ns, there are no skyrmions under the detector and the device has obtained G\textsubscript{min} value. Thus, we could achieve a linear and symmetric LTP and LTD curves for our proposed geometry of 5-bit (spiral-like K\textsubscript{u}) synapse~(Video \cite{Videolink}).
\subsection{\label{res syn conc}Synapse based on concentric Ku profile}
\vspace{-4mm}
Here, the initiation and reset time of the device is shorter as compared to that of the spiral-like K\textsubscript{u} profile device. This happens because the skyrmions are present near the vicinity of the detector region. For the same 5-bit (concentric-like K\textsubscript{u}) synapse~(Video \cite{Videolink}), the initiation time is 2 ns, potentiation is from 2 ns to 60 ns, and depression time is 66~ns to 158~ns. As shown in Fig.~\ref{fig.5}(d) at t=0~ns, no skyrmions are present in the detector region representing minimum conductance~ G\textsubscript{min}.\\~On a positive current pulse train application, the weights start to increase linearly (LTP) and attain the maximum at t~=~60~ns, all the skyrmions are under the MTJ region corresponding to maximum conductance~G\textsubscript{max}. Here, we also observe STP for 60~ns to 66~ns. The LTD begins with the application of negative current application, moving the skyrmions out of the constricted concentric paths as shown at t=102~ns, 16 skyrmions are out of the detector region. The applied current pulse train of J=2.5~MA/cm\textsuperscript{2} at pulse width and period of 1.86~ns and 2~ns respectively achieve the synaptic operations in a total duration of 158~ns~(pulse number 79) including LTP, LTD, and STP. \\
\indent Similarly, for a 4-bit (concentric-like K\textsubscript{u}) synapse~(Video \cite{Videolink}) a current density of J=2.5~MA/cm\textsuperscript{2} at pulse width and period of 1.8~ns and 2~ns respectively can achieve initiation, LTP, STP, LTD, and reset of the device in 2~ns, 2~ns to 32~ns, 32~ns to 34~ns, 34~ns to 72~ns and 72~ns to 80~ns respectively in a total duration of 80~ns for all synapse operations. \\
\indent For a 6-bit (concentric-like K\textsubscript{u}) synapse~(Video \cite{Videolink}) current density of J = 2.5~MA/cm\textsuperscript{2} at pulse width and period of 2.2~ns and 2.5~ns respectively can achieve initiation, LTP, STP, LTD, and reset of the device in 25~ns, 25~ns to 142.57~ns, 142.57~ns to 147.6~ns, 147.6~ns to 297.6~ns and 297.6~ns to 305~ns respectively in a total duration of 305~ns~(pulse number 122) for all synapse operations. The snapshots in Fig.~\ref{fig.5}(e) at t = 0~ns correspond to minimum conductance G\textsubscript{min} and t = 142.57~ns correspond to the maximum conductance G\textsubscript{max}. At t = 250.12~ns, LTD operation has already set in with 44 skyrmions being moved out of the detector heralding the reset position of the device.
\begin{figure}[t]
\centerline{\includegraphics[width=1\columnwidth]{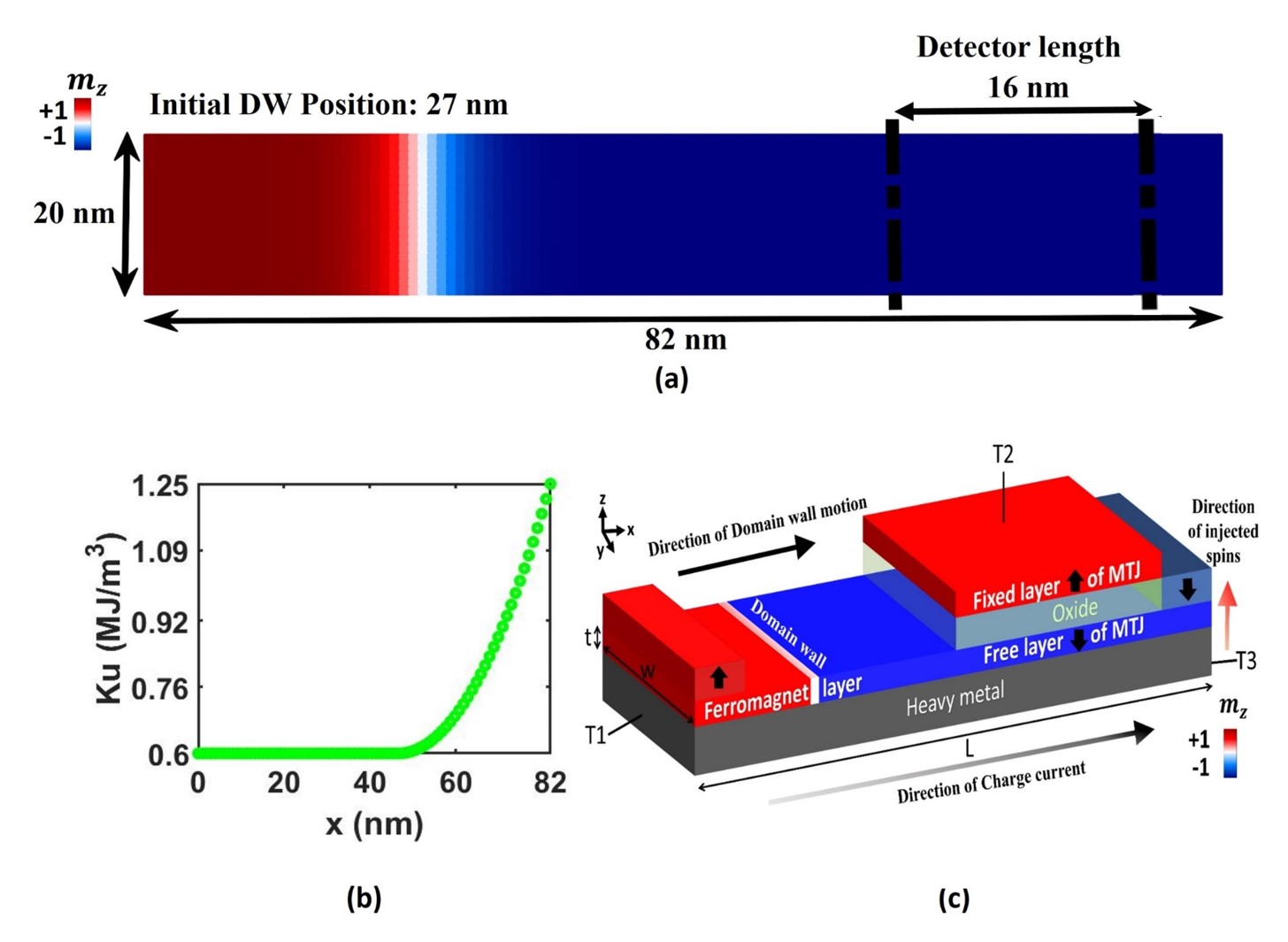}}
\caption{Domain wall-based ReLU device (a) 2D view of the domain wall track for the ReLU device (b) The applied parabolic uniaxial anisotropy~(K\textsubscript{u}) profile (PMA) along the length of the device~(x). (c)~3D view of the domain wall device in ReLU circuit}\label{Fig.3}
\end{figure}
\vspace{-0.5cm}
\subsection{\label{Domain wall-based ReLU device}Design of Domain wall-based ReLU and Max pooling device}
\begin{figure}[t]
\centerline{\includegraphics[width=1\columnwidth]{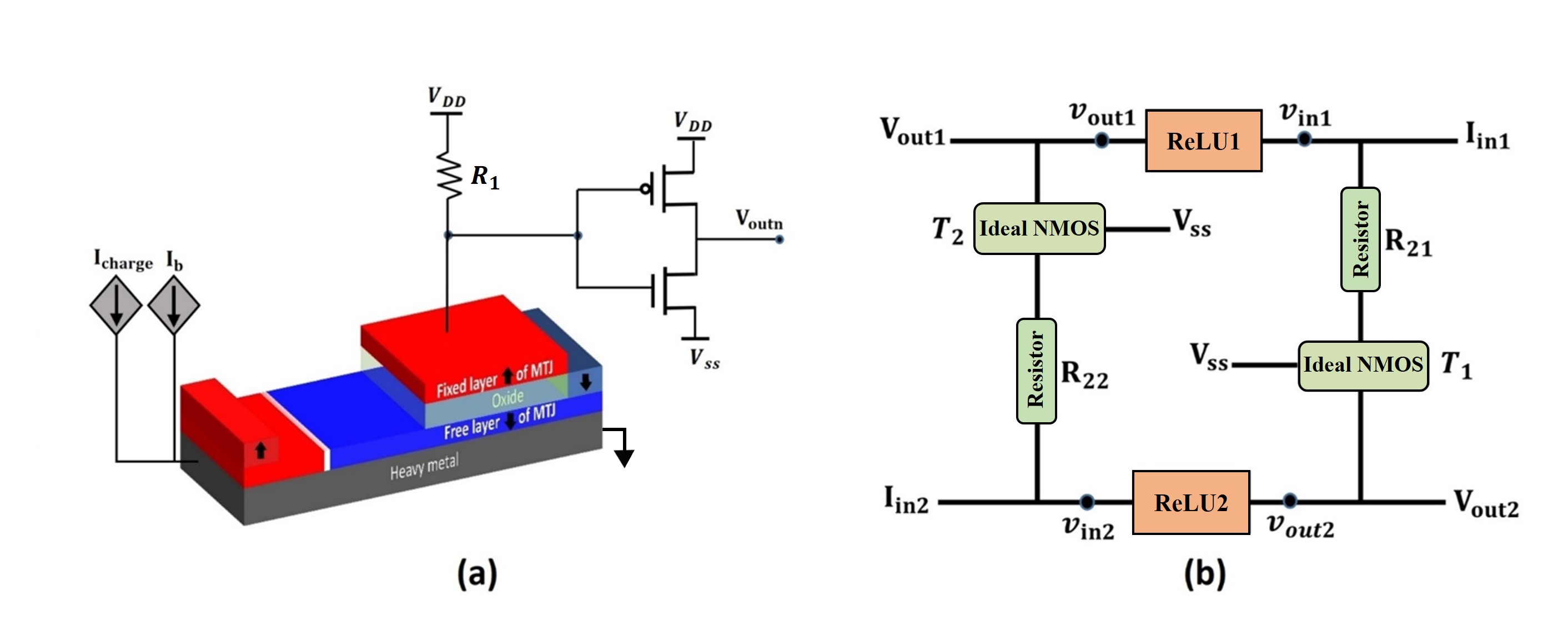}}
\caption{Domain wall-based ReLU device (a) CMOS-domain wall-based hybrid ReLU circuit (b) 2-Input ReLU Max Pooled circuit connection}
\label{Fig.4}
\end{figure}

\vspace{-4mm}
The proposed DW ReLU~(Rectified Linear Unit) device consists of a SOT-based monolayer FM~ of dimension 82~nm x 20~nm x 1~nm\cite{kumar2022domain} with a single DW (see Fig.\ref{Fig.3}(a,b)) and a CMOS-inverter circuit as shown in Fig.~\ref{Fig.4}(a). The ferromagnet layer exhibits a perpendicular magnetic anisotropy(PMA) with oppositely polarized transverse magnetic regions separated by a DW.~Figure \ref{Fig.3}(b) shows the parabolic uniaxial anisotropy profile in the FM layer of the device that can be achieved\cite{balan2023improving,fassatoui2021kinetics} using oxidation (under oxidation, oxidation, and over oxidation) of the FM layer under positive bias voltage. The Uniaxial anisotropy profile of the FM layer is constant in the left section from 0~nm to 47~nm and is parabolic in the other section from 47~nm to 82~nm. The parabolic K\textsubscript{u} allows the DW to reset at the start of the slope when no current is applied, addressing the persistent DW issue. The spin-orbit coupling at the FM-HM interface leads to the Dzyaloshinskii-Moriya exchange interaction (DMI) resulting in a stable Neel DW. The strong DMI not only rotates the DW moment but also tilts the DW line profile in the x-y axis due to the energy minimization principles reported in \cite{emori2014spin,ryu2012current}. The DW in the ferromagnet layer is driven by an electrical charge current (Terminal T1 to T3) passed through the HM layer in the x-direction giving rise to a spin-polarized current flowing in the +z-direction, having spins that are oriented along the -y-direction with a parabolic anisotropy variation under the detector region. 
For the heavy metal, a gold-platinum alloy (Au\textsubscript{0.25}Pt\textsubscript{0.75}) has been chosen over the standard platinum (Pt) metal layer to enhance power efficiency at the device level\cite{zhu2018highly,vadde2023power}.\\
\indent The domain wall read-out\cite{raymenants2021nanoscale} is done using an MTJ from `read' terminal T2 to T3. A range of conductance values corresponding to the varying DW positions is obtained for the varied current range. The fixed layer with magnetization in the +z direction along with the oxide and the free layer forms the MTJ (Fig.~\ref{Fig.3}(c)). The ferromagnet layer forms the free layer of the MTJ whose magnetization is altered by introducing the DW driven by spin-orbit torque. \\
\indent The detector is strategically placed at a distance of 63~nm to 79~nm within the structure such that the DW position shows a significant monotonic stepwise increase with respect to the applied current density. This placement allows for the implementation of ReLU-like functionality by our proposed device incorporating linearity for DW positions (see Fig.~\ref{Fig.dwr}(a)) under the MTJ and non-linearity by introducing zero otherwise. A fixed magnetization layer is used at the ends of the ferromagnet layer with an antiferromagnetic layer above to create an exchange bias that pins the magnetization at the ends of the wire, ensuring DW stabilization. Thus, protecting it from getting destroyed at the edges\cite{kaushik2020comparing,liu2018synthetic}.~Spin-orbit torque (SOT) is preferred over Spin-transfer torque (STT) for DW motion due to its faster settling time and lower current requirements for the same operation\cite{zhang2017spin,chureemart2021current,avci2017fast}. \\
\indent Figure~\ref{Fig.4}(a) depicts the in-plane current I\textsubscript{charge} as the varying input that manipulates the DW positions in the FM layer. While the bias current $\mathrm{I_b}$ shifts the output such that the CMOS-DW hybrid circuit simulates the ReLU output. The corresponding resistance values obtained from the conductance measurement are connected to a resistor~(\text{R\textsubscript{1}}) and then fed through a voltage divider circuit to a CMOS inverter. The output across the voltage divider drives the CMOS inverter pair in the linear region. Thus, facilitating the ReLU output to be implemented as a ReLU module (Fig.~\ref{Fig.4}(a)) for both activation and Max-pooling~(Fig.~\ref{Fig.4}(b)) functionality in the CNN.\\
\indent The CNN~(Fig.~\ref{fig:cnn_arch}) has four major layers: The convolutional layer, the ReLU layer, the pooling layer, and the fully connected layer. The convolutional layer is implemented using crossbar arrays connected via the spintronic device, such as the skyrmionic synapse device in our case~(\ref{Skyrmionic synapse device}). The output from the convolutional layer serves as an input to the activation or the ReLU layer, which is implemented using the SOT-based CMOS-DW hybrid circuit~(\ref{Domain wall-based ReLU device}).\\
\indent After the feature extraction stage, the max-pooling layer is implemented by linking the individual ReLU circuits in a specific design such that the ``winner takes all".~This computational principle stems from the interconnect and ReLU connection creating competition among the interconnected ReLU circuits such that only one of them corresponds to the maximum input, and others settle at zero. The interconnect consists of a resistor connected to an NMOS transistor whose gate terminal is connected to the output of other ReLU circuits. The output drawn from one ReLU circuit depends on the output voltage of another ReLU circuit. As a result of the competition, a negative current is injected into all other ReLU devices by the ReLU circuit with non-zero output at the end of the max pooling operation.~Figure~\ref{Fig.4}(b) shows ReLU plus max pooling function being simultaneously performed for two inputs. We performed a simultaneous ReLU plus max pooling for 9 Inputs which are summed without the boundation of identifying the circuit with the maximum input. Supplementary details of a similar circuit with ReLU plus max-pooling can be found in \cite{vadde2023orthogonal}.\\

\subsection{\label{res rel maxpool}ReLU and Max Pooling circuit}
\vspace{-4mm}
As described in \ref{Device Model}D~and \ref{Simulation methods}, the DW-based ReLU device is implemented. The DW, initially placed at 27~nm is relaxed and supplied with a varying electrical charge current in the range -34.8~$\mu$A to 34.8~$\mu$A. As the SOT-driven DW moves along the nano track it encounters an increasing parabolic uniaxial anisotropy profile (K\textsubscript{u} in the range 0.6~$\mathrm{MJ/m^3}$ to 1.25~$\mathrm{MJ/m^3}$) which gradually slows down the DW velocity, making the velocity zero after some time. This operation allows us to implement the ReLU-like functionality in the device utilizing the linear variation of the conductance obtained under the MTJ for DW positions at a particular instant.\\
\begin{figure}[t]
\centerline{\includegraphics[width=1\columnwidth]{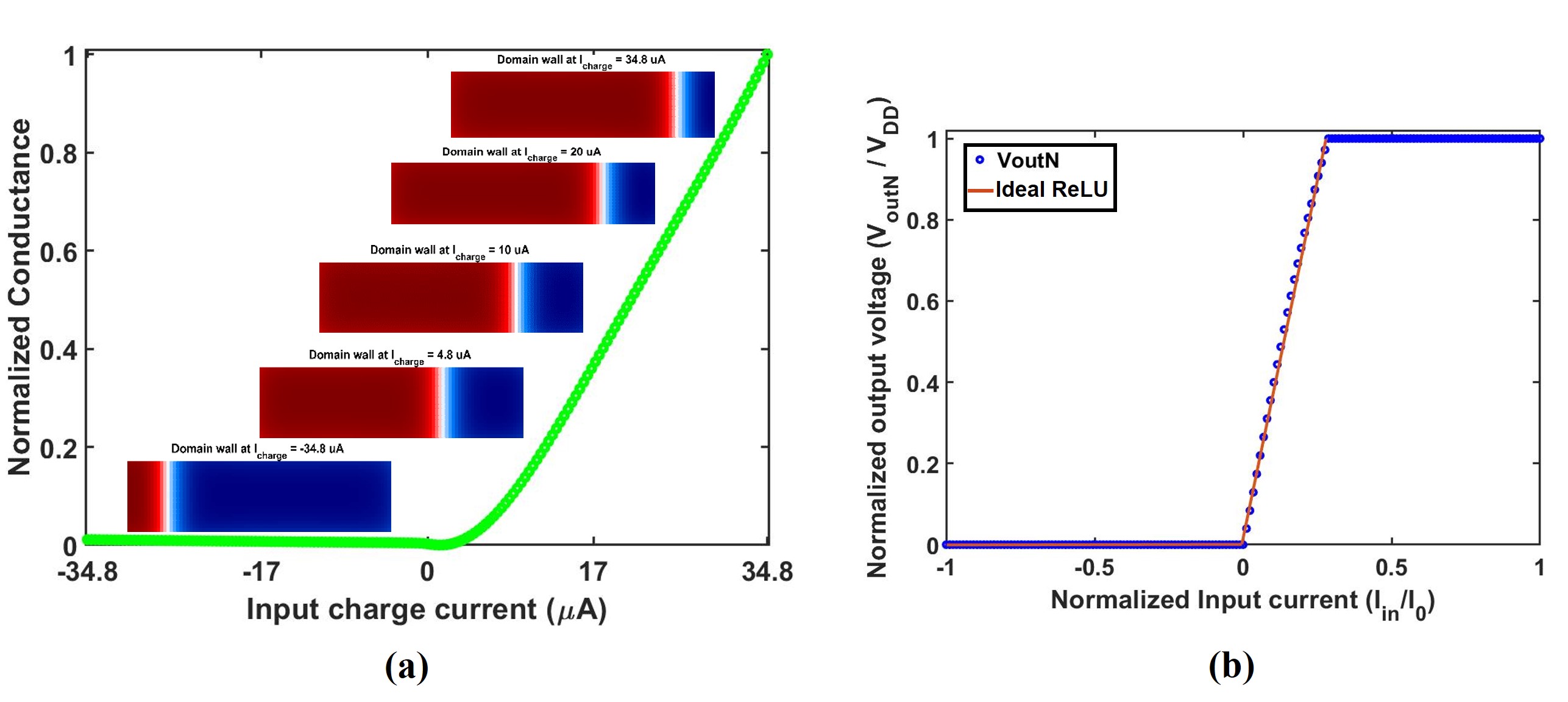}}
\caption{Domain wall-based ReLU device (a)~Normalized conductance as a function of input charge current, along with domain wall snapshots at varying input charge current (b)~ ReLU function realized using a hybrid CMOS-domain wall device.}
\label{Fig.dwr}
\end{figure}
\indent Figure~\ref{Fig.dwr}(a) shows snapshots of domain wall positions on the FM nano track. The Domain wall reaches at position = 9~nm (56.7~nm) corresponding to an un-normalized conductance value of 3.91~$\mathrm{\mu\mho}~$(4.83~$\mathrm{\mu \mho}$) at I\textsubscript{charge} = -38.5~$\mathrm{\mu A}$ (4.8~$\mu A$), ~K\textsubscript{u} = $\mathrm{0.60~MJ/m^3}$~($\mathrm{0.653~MJ/m^3}$) in 0.3~ns~(0.7~ns). The conductance here is near to the G\textsubscript{min} value~(3.125~$\mathrm{\mu \mho}$), thus the graph is flat corresponding to the non-linearity part of the ReLU function that returns zero value for any negative value. The linearity relation is observed after the DW reaches beyond 63~nm under the detector region. The DW reaches 63.78~nm with an un-normalized conductance value of 21.74~$\mu \mho$ at I\textsubscript{charge} = 14~$\mu A$, K\textsubscript{u} = $\mathrm{0.753~MJ/m^3}$ in 0.3~ns. Similarly, for I\textsubscript{charge} = 20~$\mu A$~(34.8~$\mu A$) the DW reaches 65.8~nm~(71.88~nm) with an un-normalized conductance value of 36.36~$\mu \mho$~(73.65~$\mu \mho$) at K\textsubscript{u} = $\mathrm{0.7915~MJ/m^3}$~($\mathrm{0.9316~MJ/m^3}$)  in 0.3~ns~(0.2~ns).~The $\mathrm{G_{max}( = 73.65~\mu \mho}$) is obtained at $\mathrm{I_{charge} = 33.33~\mu A}$ after which the DW position is saturated due to the pinning of the edges which restricts the DW motion in the device between 3~nm to 79~nm. \\
\indent If the DW pinning is not executed the DW will vanish beyond I\textsubscript{charge} = 36~$\mu A$. The SOT-driven DW device enables faster device reset in 1.3~ns and the DW settles in 0.2~ns to 0.3~ns range. The obtained ReLU-like function is adjusted and aligned with a biasing current (I\textsubscript{b}).
The obtained conductance values are used as a variable resistor in the voltage divider circuit as shown in Fig.~\ref{Fig.4}(a). \\
\indent Using HSPICE, embedding the information in Verilog A, an electrical read-out circuit is designed, and the variable resistor and a constant resistor (58.954~K$\Omega$) form a voltage divider feeding a CMOS inverter pair. The output obtained from the inverter circuit forms the ReLU function. The obtained ReLU functionality as shown in Fig.~\ref{Fig.dwr}(b) obtained from the hybrid CMOS DW device shows ReLU-like functionality in the range -10.67~$\mu A$ to 10.67~$\mu A$. The power consumed by the single ReLU module is 4.73~$\mu$W.
A similar graph is also obtained after the max pooling circuit.\\
\begin{figure}[t]
\centerline{\includegraphics[height=2.1in,width=2.5in]{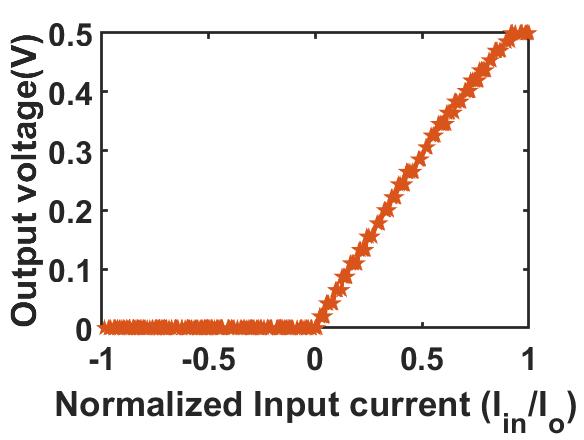}}
\caption{Normalized output voltage versus normalized input current of the Max Pooled ReLU circuit with $I_0$ = 10.67 $\mu$A.}
\label{Fig.max}
\end{figure}
\indent A specimen circuit of 2-Input ReLU-Max pooling is shown in Fig.~\ref{Fig.4}(b). An interconnect network with an ideal NMOS and a resistor (R\textsubscript{21} = R\textsubscript{22}) is attached to the ReLU circuit creating competition between the ReLU circuits as the input current drawn from ReLU1 depends on the output voltage of ReLU2.~The gate of the NMOS transistor connects to both the ReLU’s output. This design strategy creates competition calculating both the ReLU and max pooling function simultaneously. Figure~\ref{Fig.max} shows the transient response for a 9-Input ReLU Max pooling network corresponding to a 3x3 pooling layer for 400 Monte Carlo simulations consuming power of 42.42~$\mu$W.
\begin{figure}[b] 
\centering
\includegraphics[width=1\linewidth]{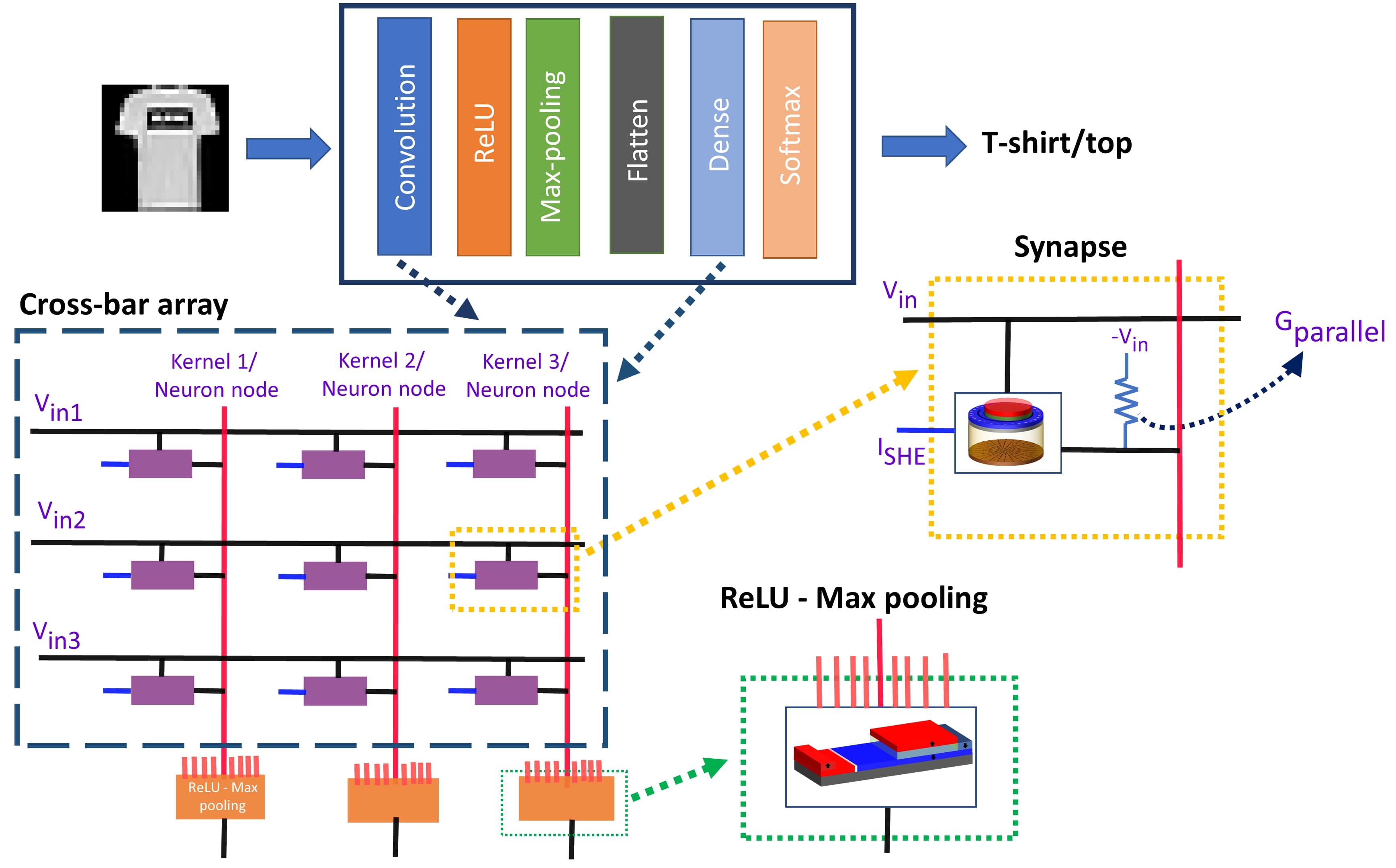}
\caption{CNN architecture for classifying fashion-MNIST and MNIST handwritten digits datasets.} 
\label{fig:cnn_arch}
\end{figure}
\subsection{\label{app CNN}Convolutional Neural Network}
\begin{figure}[t] 
\centering
    \subfigure[\label{fig:hand_written}]{\includegraphics[width=0.48\linewidth]{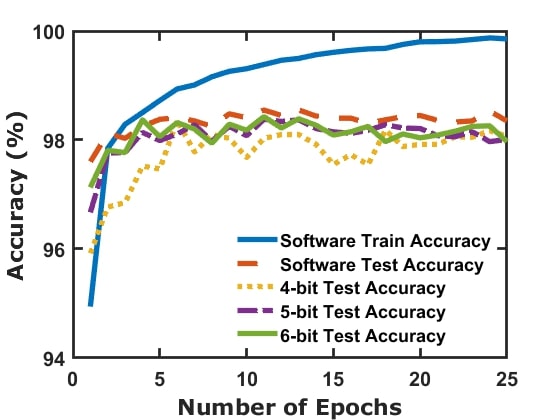}}
    \subfigure[\label{fig:fashion_mnist}]{\includegraphics[width=0.48\linewidth]{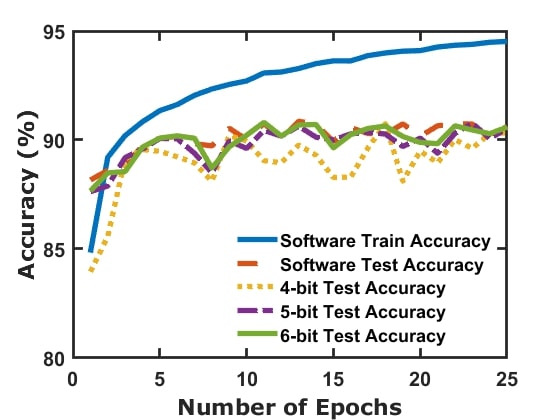}}
\caption{Classification accuracy of the CNN for software training and testing along with the inference accuracy using 4-bit, 5-bit, and 6-bit synapses~(concentric K\textsubscript{u} profiles) for both (a) MNIST handwritten digits dataset and (b) Fashion MNIST dataset.} 
\end{figure}

\vspace{-4mm}
\indent The CNNs are highly valuable in various machine learning tasks related to images, such as computer vision and image recognition \cite{goodfellow2016deep}.~Figure~\ref{fig:cnn_arch} shows the convolution and dense layers implemented using a cross-bar structure, where the input is applied to the horizontal lines. The vertical lines represent the kernel in convolution operation and neuron nodes in a fully-connected dense layer. The skyrmion device along with the parallel conductance is used to store the weight in the cross-bar array. The weights in the neural network can take both negative and positive values, whereas conductance values are strictly positive. To tackle this, we incorporate a parallel conductance (G\textsubscript{parallel}) with a negative input, allowing the synapses to effectively represent weights as shown in Fig.~\ref{fig:cnn_arch}.
\begin{equation}
   W_{i,j}=G_{skyrmion(i,j)}-G_{parallel},
\end{equation}
\begin{equation}
   G_{parallel} =\frac{G_{min} + G_{max}}{2},
\end{equation}
where, $\mathrm{W_{i,j}}$ is the weight connecting i\textsuperscript{th} row/input with j\textsuperscript{th} kernel/node, G\textsubscript{skryrmion} is the conductance of the skyrmion device and G\textsubscript{parallel} is the conductance of the device parallel to the skyrmion device. G\textsubscript{min} and G\textsubscript{max} are the minimum and maximum conductance values of the skyrmion device. The CNN training process is designed to ensure that the weights remain within the $\pm 1$ range.\\
This convolution crossbar array is then connected to the domain-wall based ReLU-max pooling circuit to complete the feature extraction stage of the CNN. The CNN architecture used for classifying Modified National Institute of Standards and Technology (MNIST) handwritten digits and fashion-MNIST datasets is illustrated in Fig.~\ref{fig:cnn_arch}. The accuracy for MNIST handwritten digits and fashion-MNIST datasets is depicted in Fig.~\ref{fig:hand_written} and \ref{fig:fashion_mnist}, respectively. The accuracies 
are presented in Table \ref{table:cnn}, demonstrating that the inference accuracies closely align with the software testing accuracy, showcasing effectiveness of our skyrmionic synapse and hybrid CMOS DW ReLU-max pooling devices. Our results indicate that the 4-bit synapse achieves an accuracy of 90.33\%, close to the 90.59\% accuracy obtained with the 6-bit implementation for the fashion-MNIST dataset. The results demonstrate an effective inference result even with the 4-bit synapse design. The power consumed by the network to process one image
~is 110 mW, consumed by the ReLU-Max pooling circuits.
\begin{table}
\centering
\caption{CNN accuracy}
 \begin{ruledtabular}
      \begin{tabular}{cm{2.5cm}m{2cm}}
\textbf{Dataset } & \textbf{MNIST Handwritten digits dataset ($\%$)} & \textbf{fashion-MNIST dataset ($\%$)}\\
\hline
Software Training & 99.85 & 94.52 \\
Software Testing & 98.35 & 90.63\\
4-bit Inference & 98.07 & 90.33 \\
5-bit Inference & 98.01 & 90.52 \\
6-bit Inference & 97.97 & 90.59\\
 \end{tabular}
\end{ruledtabular}
\label{table:cnn}
\end{table}
\section{\label{conc}Conclusion}
\vspace{-4mm}
 In summary, the paper presents a multi-state
 skyrmion based synapse device designs of 4-bit(16 states), 5-bit(32 states) and 6-bit(64 states) 
 with skyrmions as neurotransmitters driven by the current-induced torque. The two kinds of labyrinth-maze like uniaxial anisotropy profiles introduced in the circular ferromagnet disc enable constricted and aligned skyrmion motion facilitating weight change as they gyrate under the detector, read using magnetoresistance. The synapse device achieves a highly desired symmetric and linear weight update during LTP and LTD operations and an ultra-low energy consumption of 0.87~fJ without additional techniques, overcoming the common challenge of nonlinear weight update\cite{boybat2018neuromorphic,lee2022crus}. Additionally, we have designed a 9-Input ReLU-Max pooled circuit with a power consumption of 42.42~$\mu$W using a hybrid CMOS-SOT-driven domain wall device consistent with the cross-bar array using a skyrmion-based synapse for implementing the convolutional layer.~Our comprehensive CNN model is evaluated for pattern recognition using the MNIST handwritten~(fashion MNIST) datasets, achieving an accuracy of approximately 98.07\%(~90.33\%), comparable to software-based training accuracy of around 98.35\%(~90.63\%) for 4-bit synapse design and a power consumption of 110 mW to process one image in the network. Our work extends the scope for designing a maximal hardware-based neuromorphic computing platform employing spintronic devices by coupling an atomistic description to circuit modeling.
\vspace{-4mm}
\section{\label{Simulation methods}Methods}
\begin{figure}
\centerline{\includegraphics[width=1\columnwidth]{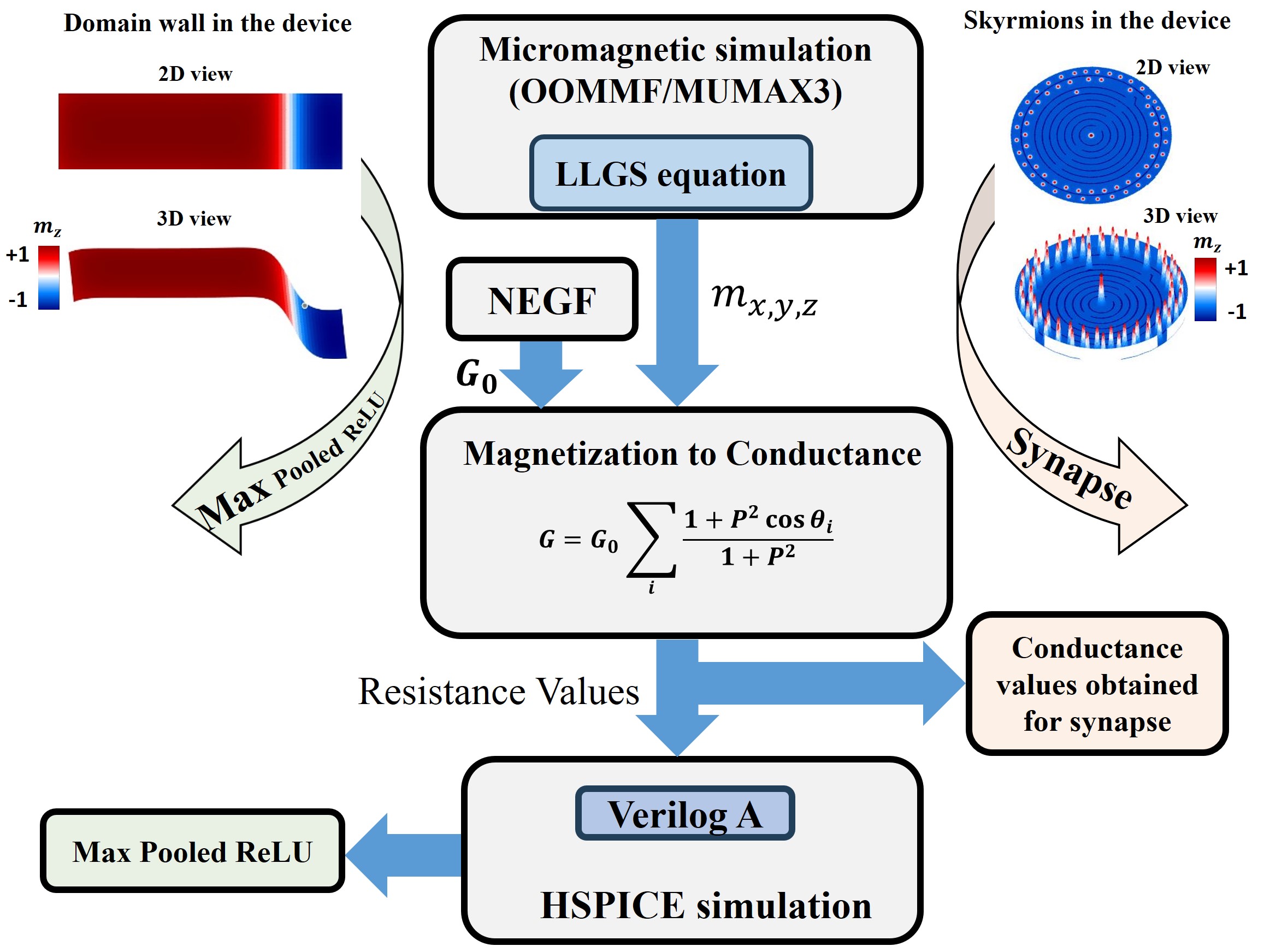}}
\caption{Simulation methodology}
\label{Fig.sim}
\end{figure}

\vspace{-4mm}
\indent The micro-magnetic simulations for the synapse are performed using OOMMF\cite{donahue1999oommf} with the added DMI extension module. The dimensions of the synapse devices are mentioned in Table~I with discretization of 2 nm x 2 nm x 0.5 nm. The proposed synapse devices are simulated using the parameters mentioned in Table \ref{table2} for the Co-Pt system\cite{sampaio2013nucleation}.
The micro-magnetic simulations for the DW-based ReLU are performed on a GPU-accelerated numerical package (mumax3). The DW device has a thickness of 1nm with lateral dimensions 82~nm x 20~nm. The cell size is 1 nm x 1 nm x 1 nm. The proposed DW is simulated using the matching parameters mentioned in Table \ref{table3} for the $\mathrm{Co-Au_{25}Pt_{75}}$ system\cite{yang2023magnetic}. The uniaxial anisotropy for the device length 0~nm to 47~nm remains constant at $\mathrm{Ku_c = 0.6~MJ/m^3}$. The profile varies parabolically for the device length 47nm to 82nm with the relation 
 $\mathrm{Ku= Ku_c + 530.6122*(i-47)^2}$ where i varies from 47 to 82 representing distinct regions, as per the mumax3 code.\\
\indent The hybrid device circuit for ReLU-Max pooling is simulated in HSPICE using Verilog A. The proposed circuit parameters are mentioned in Table \ref{table3}.\\
\indent A schematic overview of the simulation methodology is shown in Fig.~\ref{Fig.sim}, A detailed explanation for both skyrmion and DW devices is mentioned below in subsections A, B, C, D, and E.
\subsection{\label{oommfskyrmion}Micromagnetic simulation framework for the skyrmion-based synapse device}
\vspace{-4mm}
The magnetization dynamics for CPP are computed using the Landau-Lifshitz-Gilbert (LLG) equation with the Slonczewski spin transfer torque(STT) is defined as:
\begin{eqnarray}		
	\frac{d\textbf{m}}{dt}&=&-|\gamma|\textbf{m}\times\textbf{H}_{eff}+
	\alpha \left( \textbf{m}\times \frac{d\textbf{m}}{dt}\right)\nonumber \\
	&&+ |\gamma|\beta_j \epsilon \left( \textbf{m}\times \textbf{m}_{RL}\times \textbf{m}\right) -|\gamma|\beta_j \epsilon^{\prime}\left(\textbf{m}\times\textbf{m}_{RL}\right) 	\nonumber,\\		
\end{eqnarray}
where, $\gamma=2.211\times10^5~ \text{m/(A.s)}$ is the gyromagnetic ratio, $\textbf{m}_{RL}$ is the magnetization of RL, $\beta_j=\hbar J/(\mu_0 e t_z M_s)$ , 
 $\epsilon=\frac{P\Lambda^2}{\left(\Lambda^2+1\right)+\left(\Lambda^2-1\right)\cdot\left(\mathbf{m}\cdot\mathbf{m}_{RL}\right)}$, and $\epsilon^{\prime}$ is the secondary spin transfer term.we have set $\Lambda$=1, and the secondary spin transfer term $\epsilon^{\prime}$=0 to remove the dependence of $\epsilon$ on $m.m_{RL}$ and the field like out of plane torque respectively in all simulations. 
The $\hbar$, e, t, P, and J are reduced Planck's constant, electronic charge, thickness of the FM layer, polarization, and current density, respectively. The effective magnetic field $\textbf{H}_{eff}$ in relation to local magnetization is given by:
\begin{equation}\label{Hfield}
	\textbf{H}_{eff}=-\frac{1}{\mu_0 M_s} \frac{\delta E}{\delta \textbf{m}},
\end{equation}
where $\mu_{0}$ is the space permeability and E is the total energy given as follows:
\begin{eqnarray}\label{energy}		
	E&=&\int dV \big[ \big. A(\nabla \textbf{m})^2 + \varepsilon_{DMI} + K_u\left[1-(\textbf{m}\cdot\textbf{z})^2\right]\nonumber \\
	&& - \frac{\mu_0}{2} \textbf{m} \cdot \textbf{H}_d],		
\end{eqnarray}	
where the first term is the exchange energy with the exchange stiffness constant A, the second term is an interfacial form of DMI resulting in a Neel skyrmion $\varepsilon_{DMI}=D\left( m_z\frac{\partial m_x}{\partial x}-m_x\frac{\partial m_z}{\partial x}+m_z\frac{\partial m_y}{\partial y}-m_y\frac{\partial m_z}{\partial y}\right)$. The third term defines the uniaxial anisotropy energy with the anisotropy constant K\textsubscript{u}, and the demagnetization field forms the last term.
\begin{table}[b]
\caption{The micromagnetic simulation parameters for the synapse device}
\begin{ruledtabular}
  \begin{tabular}{ccc}
  \textbf{Symbol} & \textbf{Parameters} & \textbf{value}\\ \hline
$M_{s}$ & Saturation magnetization & 580KA/m \\
$\alpha$ & Gilbert damping factor & 0.3 \\
$D$ & DMI constant & 3mJ/$m^{2}$ \\
$A_{intra}$ & Exchange stiffness constant & 15pJ/m\\
$P$& Spin polarization factor & 0.4 \\
$K_{u}$ & PMA constant & 0.8MJ/$m^{3}$\\

\end{tabular}
\label{table2}
\end{ruledtabular}
\end{table}
\subsection{\label{skyrmion dynamics}Skyrmion dynamics}
\vspace{-4mm}
\indent The Skyrmion dynamics in the free FM layer controlled by the spin current is described by the Thiele equation :
\begin{equation}
	\textbf{G} \times \textbf{v} -\alpha \mu_0 M_s t_z \textbf{v}.\textbf{d}/\gamma + \textbf{F}_{STT} + \textbf{F} = 0.
	\label{thiele_eq}
\end{equation}

The first term is the Magnus force with the velocity of skyrmion and the gyrovector $\textbf{G}$. The second term describes the dissipative force with the damping $\alpha$ and the dissipative tensor $\textbf{d}=\begin{pmatrix}
d & 0\\
0 & d
\end{pmatrix}$, where $d= \frac{1}{4\pi}\int dx dy (\partial_x \textbf{m}\cdot\partial_x \textbf{m})$.
The third term represents current-induced driving force due to the spin transfer torque, which can be decomposed into two orthogonal directions~(the radial $\hat{r}$ and the tangential direction $\hat{t}$ in the polar coordinates ),~$\textbf{F}_{STT}=F_r \hat{r} + F_t \hat{t}$, the component $F_{i}$ where (i = r or t) is given by $F_{i}=-\mu_0 \beta_j M_s t_z \int dx dy[(\textbf{m}\times\textbf{m}_p) \cdot \partial_{i}\textbf{m}]$, where $\beta_{j}$ is the applied vertical current density. The last term is the boundary-induced force which includes edge repulsion, the nearest neighbor skyrmion repulsion forces, and little effect of the high uniaxial anisotropy outer ring ( only in the case of 6-bit skyrmionic synapse) when the skyrmion moves steadily in the nanodisk~(means radial part of the velocity is zero). Therefore, Eq.~\ref{thiele_eq} can be written as :
\begin{eqnarray}\label{thiele_2a}
    v_tG+F_r+F=0, \\ 
    \label{thiele_2b}
    \alpha \mu_0 M_s t_z v_td/\gamma + F_t=0.
\end{eqnarray}
From the above two equations, we deduce that the steady motion of skyrmion in the nanodisk requires a balance between the Magnus force and the boundary force. Thus, the skyrmions will move towards the center of the nanodisk at the positive current pulse train and away from the center on the application of a negative pulse train.

\subsection{\label{mumax dw}Micromagnetic simulation framework for Domain wall-based ReLU device}
\vspace{-4mm}
The methodology for micromagnetic modeling of the magnetization dynamics in spin-orbit-torque driven DW is reported in Ref. \cite{joos2023tutorial}. The GPU-accelerated micromagnetic package mumax3 uses custom field functionality to implement spin-orbit torque (SOT) with the Landau-Lifshitz-Gilbert (LLG) equation given as:
\begin{eqnarray}\label{LLG}		
	\frac{d\textbf{m}}{dt}&=&-|\gamma|\textbf{m}\times\left( \textbf{H}_{eff}+ \textbf{H}_{ST}\right)+\alpha \left( \textbf{m}\times \frac{d\textbf{m}}{dt}\right)\nonumber, \\		
\end{eqnarray}
where $\mathbf{H}_{\mathrm{eff}}$ is the effective field and $\mathbf{H}_{\mathrm{ST}}$
 is the spin torque consisting of the damping-like and the field-like term. 
\begin{equation}\label{hsot}
	\textbf{H}_{ST} =a_J\textbf{m}\times\textbf{p}+b_J\textbf{p},
\end{equation}
\begin{table}[b]
\caption{The simulation parameters for domain wall-based ReLU Max Pooled device}
 \begin{ruledtabular}
      \begin{tabular}{ccc}
      \textbf{Symbol} & \textbf{Parameters} & \textbf{value}\\ \hline
      $M_{s}$ & Saturation magnetization & 580KA/m \\
      $\alpha$ & Gilbert damping factor & 0.3 \\
      $D$ & DMI constant & 3mJ/$m^{2}$ \\
      $A_{intra}$ & Exchange stiffness constant & 15pJ/m\\
      $P$ & Spin polarization factor & 0.15 \\\\
      $\xi$ & $\frac{|field-like torque|}{|damping-like torque|}$ & -2\\\\
      $\theta$ & Spin Hall Angle & 0.3\\
      $R_{HM}$ & Resistance of HM & 850.75$\Omega$\\
      $\rho_{HM}$ & resistivity of HM & 83$\mu\Omega cm$\\
     $t_{HM}$ & Thickness of HM & 4nm\\
     $L_{HM}$,$W_{HM}$& length, Width of HM & 82nm, 20nm\\
     L,W & Length,Width & 82nm, 20nm \\
     V & Volume of FM & 1640$nm^3$\\
     $C_{MTJ}$ & MTJ Capacitance & 26.562aF \\
     $R_1$ & Reference Resistor & 58.954K$\Omega$ \\
     $R_{21}$ & Resistor in NMOS current source & 42K$\Omega$\\
     $I_b$ & Biasing current & 39.735uA \\
     $V_{DD}$, $V_{SS}$ & Voltage sources & 0.5V, -0.5V\\
    $\Delta_t$ & Simulation time step & 0.05ps\\
\end{tabular}
\end{ruledtabular}
\label{table3}
\end{table}
where the unit vector p characterizes the direction of the spin-polarization.The $a_J$, $b_J$, and $\xi$ ( $\xi$=$b_J/a_J$)are the magnitudes of the damping-like, field-like terms and the ratio of the field-like term and damping-like term respectively. 
The spin current from the heavy metal layer is given by Hirsch\cite{hirsch1999spin}, Takahashi, and Maekawa\cite{takahashi2008spin}
\begin{equation}
	I_s =\theta\frac{l_{FM}}{t_{HM}}I_c \times P
	\label{spin current},
\end{equation}
where $\mathrm{I_s, \theta, l_{FM}, t_{HM}, I_c}$, and P are the magnitude of spin current, spin Hall angle, length of the FM, thickness of the HM layer, charge current and polarization of the spin current respectively. A charge current in the +x direction injects a y-polarised spin current in the FM layer in the z-direction, inducing DW motion along the device length.
The resistance of the HM is given by $\mathrm{R=\rho l_{HM}/W_{HM}t_{HM}}$ , where $\rho$ is the resistivity of HM~($\mathrm{Au_{0.25}Pt_{0.75}}$), $\mathrm{l_{HM}}$ is the length of HM, $\mathrm{W_{HM}}$ is the width of HM and $\mathrm{t_{HM}}$is the thickness of HM.
\begin{table}[t]
\caption{The parameters for energy consumption per weight update for our proposed synapse device}
 \begin{ruledtabular}
      \begin{tabular}{cm{5cm}c}
      \textbf{Symbol} & \textbf{Parameters} & \textbf{value}\\ \hline
      \\
      $\rho_{Co}$ & Resistivity of FM(Co) & 19~$\mu\Omega cm$ \cite{milosevic2018resistivity} \\\\
      $\rho_{Pt}$ & Resistivity of NM(Pt) & 100~$\mu\Omega cm$ \cite{nguyen2016spin}\\\\
      $\rho_{Ru}$ & Resistivity of NM(Ru) & 18~$\mu\Omega cm$ \cite{wen2016ruthenium} \\\\
      $t_{Co}$ & Thickness of FM(Co) & 0.5~nm\\\\
      $t_{Pt}$ & Thickness of $NM_{1}$(Pt) & 0.5~nm \\\\
      $t_{Ru}$ & Thickness of $NM_2$(Ru)& 7~nm \cite{wen2016ruthenium}\\\\
      $l_{sf}^{Co}$ & Spin flip length of FM & 60~nm \cite{nguyen2014spin}\\\\
      $l_{sf}^{Pt}$ & Spin flip length of $NM_1$ & 0.5`nm-14~nm \cite{nguyen2014spin}\\\\
     $l_{sf}^{Ru}$ & Spin flip length of $NM_2$ & 14~nm \cite{khasawneh2011spin}\\\\
     $r_b^{Co/Pt}$ & Interface resistance of FM/$NM_1$ &  1.7$\pm$0.25~f$\Omega$$m^2$ \cite{sharma2007specific}\\\\
     $r_b^{Co/Ru}$ & Interface resistance of FM/$NM_2$ & 0.5 ~f$\Omega$$m^2$ \cite{khasawneh2011spin}\\\\
     $\beta_{Co}$ & Bulk spin asymmetry coefficient of FM& 0.46 \cite{sharma2007specific}\\\\
     $\Gamma_{Co/Pt}$ & Interfacial spin asymmetry coefficient of $NM_1$ & 0.38$\pm$0.06 \cite{sharma2007specific} \\\
    $\Gamma_{Co/Ru}$& Interfacial spin asymmetry coefficient of $NM_2$ & -0.2 \cite{khasawneh2011spin}
 \end{tabular}
\end{ruledtabular}
\label{table4}
\end{table}
\subsection{\label{electrical readout}Electrical read out}
\vspace{-4mm}
When all the skyrmions~(domain wall) reach the detector region, the TMR calculation is done by a differential method, considering that the whole magnetic tunnel junction (MTJ)~device consists of many 2 nm × 2 nm (1 nm x 1 nm) uniform cells in the xy plane \cite{sawada2009tunneling}. The conductance of each cell is obtained by the extended Julliere’s model\cite{tanaka2001large,julliere1975tunneling} and slonczewski conductance model\cite{slonczewski1989conductance} which can be shown as:
\begin{equation}\
\label{conductance}
 G = G_0 \displaystyle\sum_{i}\textbf( \frac{ 1+ P^{2}  cos \theta}{1+ P^{2} }).
\end{equation}
NEGF is used to calculate the $\mathrm{G_{0}}$, where $\mathrm{G_{0}}$ is the conductance when the magnetization is perfectly parallel to the reference layer, P is the spin polarization which is set to be 0.4 in our simulation, and $\mathrm{\theta}$ is the magnetization of each cell with respect to the reference layer. 

The conductance obtained is then converted into weights on a normalized scale using the relation:
\begin{equation}
\label{weight}
Weight = \frac{G - G_{min}}{G_{max} -G_{min}},\\
\end{equation}
here, G\textsubscript{max}~(G\textsubscript{min}) is the maximum (minimum) conductance of the synapse.
We further calculated the write resistance $\mathrm{R_{Write}}$ of our proposed device using the Valet and Fert model~\cite{fert1995spin,valet1993theory} for a bilayer structure consisting of ferromagnet layers~(resistivity~($\mathrm{\rho_F^*}$), thickness~ ($\mathrm{t_F}$),~$\beta_F\neq 0$) and the non-magnetic layer~ ($\mathrm{\rho_{N1}^*,~\rho_{N2}^*,~t_{N1},~t_{N2}}$) under the case of spin-flip length $l_{sf}^{F, N1,N2} >> t_F, t_{N1},t_{N2}$:
\begin{equation}\label{AreaxResistance}
       Area * R_{write} = r_{0} + r_{SI}(\theta),\\
\end{equation}
\begin{equation} \label{rtheta}
      r_{SI}(\theta) = r_{SI}^P + ( r_{SI}^{AP} - r_{SI}^P)\frac{(1-\cos(\theta))}{2},\\
\end{equation}
\begin{equation}\label{r(0)}
       r_{0} = 2\rho_{f}^* t_F + \rho_{N1}^* t_{N1} + \rho_{N2}^* t_{N2},\\
\end{equation}
\begin{equation}\label{r_SI_AP}
      r_{SI}^{AP} = 2\rho_{f}^* t_F + \rho_{N1}^* t_{N1} + \rho_{N2}^* t_{N2} + 2r_b^{F/N1} + 2r_b^{F/N2},\\
\end{equation}
\begin{eqnarray*}\label{r_SI_P}
      \left(r_{SI}\right)^P & = &  r_{SI}^{AP} - (\beta_{F}\rho_{f}^*\frac{t_F}{t_F+t_{N1}} L_{1}\\
      & & + \beta_{F}\rho_{f}^*\frac{t_F}{t_F+t_{N2}}L_{2}\\
      & & + 2r_b^{F/N1} \Gamma^{F/N1} + 2r_b^{F/N2}\Gamma^{F/N2})^2)/ r_{SI}^{AP},
\end{eqnarray*}
Where, r\textsubscript{0} is the resistance of the nonmagnetic layer in series with a magnetic one,~r\textsubscript{SI}($\theta$) is the spin-coupled interfacial resistance accounting for the angular dependence\cite{rijks1993combined} of the RL and free FM layer with $\theta$ = 90\textsuperscript{o} in Eq.~\ref{AreaxResistance} and Eq.~\ref{rtheta}.
The spin-coupled interfacial resistance for Parallel (Anti-parallel)~({r\textsubscript{SI}\textsuperscript{P,AP}}) configuration is defined in Eq.~\ref{r_SI_AP} and Eq.~\ref{r_SI_P}.~The resistivity of the Ferromagnet and non-magnetic layer is $\rho$\textsubscript{F}\textsuperscript{*}= $\rho$\textsubscript{F}/(1-$\beta$\textsubscript{F}\textsuperscript{2}) and~$\rho$\textsubscript{N}\textsuperscript{*} = $\rho$\textsubscript{N} respectively.~The $\mathrm{r_b^{F/N1} and~r_b^{F/N2}}$ are the interface resistances between the Co/Pt and Ru/Co.~The $\mathrm{L_i=t_F + t_{N_i}}$, where i=1, 2.~While the bulk and interfacial spin asymmetry coefficient is defined as $\mathrm{\beta_F}$ and $\mathrm{\Gamma^{F/N1, F/N2}}$.\\
\indent The parameter values used for the resistance calculation are mentioned in \ref{table4}.
The energy consumption required to move the skyrmions in the nano tracks for hardware implementation in the neural network is calculated using:
 \begin{equation}
\label{Eenergy}
 E_{write} = I_c^2 R_{wite}T\textsubscript{p},
\end{equation}
here, $\mathrm{E_{write}}$ is the Energy dissipated per weight update,~$\mathrm{I_{c}}$ is the charge current, and T\textsubscript{p}~(pulse period) is the length of write pulse required to change the weight of our synapse device to move each skyrmion from the pre-synaptic region to the post-synaptic region to calculate the energy consumption per weight update.
\vspace{-7mm}
\subsection{\label{hspice using verilogA}HSPICE using Verilog A}
\vspace{-4mm}
The obtained resistance values from the electrical readout as described in \ref{electrical readout} are utilized in the circuit-level simulation of the ReLU max-pooling circuit using Verilog A. In the HSPICE simulation of a single ReLU, the obtained resistance values act as a variable resistor connected to a constant resistor~(R\textsubscript{1}) of value 58.954~K$\Omega$, creating a voltage divider. The output from the divider is fed as an input to the CMOS inverter pair based on the 16nm node of the predictive technology model~(PTM).
A 9-input interconnection of such ReLU circuits with an interconnect network consisting of resistors and NMOS transistors is implemented in HSPICE to obtain the combined result of the ReLU-max pooling function.
\vspace{-4mm}
 \section*{\textbf{Data availability}}\vspace{-4mm}The datasets generated during and/or analyzed during the current study are available from the corresponding author on reasonable request.
 \vspace{-0.5mm}
 \section*{\textbf{Code availability}}\vspace{-4mm}The codes that support the findings of this study are available from the corresponding author upon reasonable request.
\bibliography{reference}
\vspace{-4mm}
\section*{Acknowledgements}\vspace{-4mm}
The author BM wishes to acknowledge the support by the Science and Engineering Research Board (SERB), Government of India, Grant No. MTR/2021/000388, and the Ministry of Education, Government of India, via Grant No. STARS/APR2019/NS/226/FS under the STARS scheme. The author SG acknowledges the financial support from the University Grants Commission (UGC). The author, AS, acknowledges the support provided by SERB, Grant No. SRG/2023/001327.\\
\vspace{-8mm}
\section*{Author contributions}\vspace{-4mm}
A.S. conceived the core idea of this study. S.G. performed the micromagnetic, and circuit simulation. V.V.performed the CNN implementation. S.G., V.V., B.M. and A.S. prepared the manuscript. All authors contributed to the
interpretation of the results and discussions.\\\\
\vspace{-1mm}
\textbf{Competing interests}
The authors declare no competing interests.\\\\
\textbf{Correspondence} and requests for materials should be addressed to
Abhishek Sharma.
\end{document}